# Large and temperature-independent piezoelectric response in $Pb(Mg_{1/3}Nb_{2/3})O_3$-$BaTiO_3$-$PbTiO_3$

Lizhu Huang[1,2], Guorong Li[1,a], Desheng Fu[3], Suzie Sheng[4], Jiangtao Zeng[1], Huarong Zeng[1]

[1]Shanghai Institute of Ceramics, Chinese Academy of Sciences, Shanghai 200050, China
[2]Graduate School of Chinese Academy of Sciences, Beijing 100039, China
[3]Shizuoka University, Johoku 3-5-1, Naka-ku, Hamamatsu 432-8561, Japan
[4]AM Chemicals, LLC, 721 Sunset Drive, Vista, CA 92081, USA

**ABSTRACT**

The temperature dependence of elastic, dielectric, and piezoelectric properties of $(65-x)[Pb(Mg_{1/3}Nb_{2/3})O_3]$–$xBaTiO_3$-$35PbTiO_3$ ceramics with x=0, 1, 2, 3, and 4 was investigated. Compound with x=2 was found to exhibit a large piezoelectric response ($d_{31}\approx$-170 pC/N, $d_{33}\approx$530 pC/N at 300 K). Particularly, its $d_{31}$ value was nearly a constant over a temperature range from 185 to 360 K. A broad ferroelectric phase transition tuned by $BaTiO_3$ doping was deduced from the dielectric constant, elastic compliance constant and Raman spectra. The temperature-stable piezoelectric response was attributed to the counter-balance of contributions from the dielectric and elastic responses.

**PACS:** 77.84.Ek, 77.80. B-, 77.65.Bn

Pb(Zr$_x$Ti$_{1-x}$)O$_3$ (PZT) and (100-x)[Pb(Mg$_{1/3}$Nb$_{2/3}$)O$_3$]-xPbTiO$_3$ (PMN-xPT) piezoelectric materials with a ABO$_3$ perovskite structure have been employed widely in actuators. For some special electromechanical devices, a large and temperature-stable piezoelectric response over a wide temperature range is required. Unfortunately, piezoelectric properties of many materials depend heavily on temperature[1-3]. It has been suggested that the reduction in piezoelectric constant is likely due to the "freezing" of the extrinsic contribution arising from the domain motion[2] and phase transition[4,5] at cryogenic temperatures. Other possible mechanism includes the low elastic compliance caused by hardening of lattice mode at low temperature. Hard doping, which reduces extrinsic contributions, is considered as an effective way to achieve temperature-stable piezoelectric properties in the case of PZT based ceramics[3]. However, such method displays a major drawback, that is the dielectric and piezoelectric activities of the doped ceramics are reduced dramatically at near ambient temperature[3].

PMN-xPT based materials have attracted considerable interests over the past decade owing to their ultrahigh piezoelectric response ($d_{33}$>1500 pC/N, $k_{33}$>95% for PMN-PT crystal) at near room temperature in comparison with that of PZT ceramics[6,7]. Recently, several monoclinic (M) phases have been discovered in the composition close to the morphotropic phase boundary (MPB) by Raman[8], synchrotron x-ray powder diffraction[9], and powder neutron diffraction[10]. According to Singh *et al*'s report[10], PMN-xPT with the composition around MPB has the following structural characteristics for 31≤x≤40: (1) in addition to the paraelectric cubic to ferroelectric tetragonal (*T*) phase transition, a *T* to $M_c$ phase transition occurs upon cooling, while the $M_c$ phase with Pm space group remains stable down to the lowest temperature measured. This is contrast to the case of x>40 for which *T* phase is a ground state; (2) the temperature range, over which the Mc and T phase coexist increases with increasing Ti$^{4+}$ content. For example, *M-T* phase transition is shown to occur between 300 and 200 K for x=36, indicating the broadening of *M-T* phase transition in PMN-PT. Here we define such phase transition as "broad phase transition". Despite of the intensive researches, the effect of this broad phase transition on the piezoelectric properties is still not fully understood.

In this work, $BaTiO_3$ doping was introduced to the PMN-PT system to fine tune the *M-T* phase transition. The effects of this low temperature broad phase transition and mechanical elastic response on piezoelectric properties were systematically investigated in the system of (65-x)PMN-xBT-35PT. More importantly, large and stable piezoelectric response has been achieved over a broad operating temperature range in this system.

The (65-x)PMN-xBT-35PT (x=0,1,2,3,4) ceramics have been prepared by conventional solid-state reaction methods. The samples have sizes of $\Phi 15\times 0.5$ mm. Dielectric, piezoelectric, and elastic properties as a function of temperature from 120 to 360 K were measured by a resonant-antiresonant method suggested by IEEE standards[11] using Agilent HP4294A impedance analyzer. The temperature with a precision of ±0.01 K was controlled by Novocontrol Quatro Cryosystem. The piezoelectric constants $d_{33}$ at room temperature were measured by the direct piezoelectric effect method using a piezo-$d_{33}$ meter (Institute of Acoustic Academia Sinica, model ZJ-4AN).

The dielectric properties at 1 kHz for the poled PMN-BT-PT ceramics heated from 120 to 360 K are shown in Fig. 1(a). Similar to PMN-PT ceramics[10], PMN-BT-PT ceramics also show a stage-like dielectric anomaly below the temperature for the para-ferroelectric phase transition ($T_c$). This dielectric anomaly may be attributed to the broad *M-T* phase transition[10]. With increasing *x,* this dielectric anomaly shifts to lower temperature accompanied by the smearing of dielectric stage, resulting in the flattening of dielectric response in a wide temperature range for composition of x=2, 3, and 4. In contrast, $T_c$ remains nearly unchanged as shown in the insert of Fig. 1(a). The dielectric loss was also showed in Fig. 1(a). All the samples show relatively small dissipation factor at room temperature. The dramatic increase on dissipation factor at low temperatures might attributed to phase transition or domain wall relaxations [1].

**Fig. 1(b)** reveals the temperature dependence of the elastic compliance constant ($s_{11}^E$). Clearly, $s_{11}^E$ shows a well-defined peak for all compositions. However, the elastic peak broadens and shifts gradually to lower temperature without substantial loss in its peak value when the amount of BT increases. It is noteworthy that the maximal elastic value for *x*=2 at ~230K remains as high as that for *x*=0 at ~330K. For normal solids, $s_{11}^E$ is expected to decrease upon cooling. The existence of elastic peak suggests the PMN-BT-PT ceramics undergo structural phase transition in this temperature region, during which the acoustic lattice mode becomes instable[12]. This explanation is also supported by the coincidence of the temperature at which the stage in dielectric constant occurs and that at which the elastic compliance peaks.

To better understand the M-T broad phase transition, Raman scattering spectroscopy was performed on the poled samples with x=1 and 2. The Raman spectra were recorded in back scattering geometry by Renishaw inVia Raman microscopes with a 514 nm argon ion laser excitation source, a 10 μm spot size of laser beam and an Oxford Variox temperature controlled systems. The obtained Raman spectra were fitted by Gaussian functions. Similar to what has been reported by Slodczyk *et al* [8] on PMN-xPT crystal, the Raman spectra do not show marked variations with temperature as shown in the insert of Fig. 2. However, a detailed qualitative analysis reveals a significant change in the intensity ratio of the scattering at 750 cm$^{-1}$ ($I_{750}$) and that at 510 cm$^{-1}$ ($I_{510}$), in the temperature range from 200 to 300 K and 260 to 300K for *x*=2 and *x*=1, respectively (Fig. 2). This phenomenon has been ascribed to the M-T broad phase transition in PMN-PT[13,14]. These two wavelengths are selected because they originate from the B-O stretching modes[15]. The marked change in intensity ratio of these two scatterings can be considered as a manifestation of the evolution of the coexisting phases during the broad phase transition since Raman scattering intensity is proportional to volume fraction of the corresponding phases with Raman active[15]. Remarkably, the temperature region where marked change in Raman scattering intensity occurs coincides with that over which observed elastic anomaly occurs. Raman analysis also shows that broad phase transition occurs in a wider temperature range with the increase of BT content.

The dielectric, elastic and Raman studies suggest that the elastic compliance and Raman intensity change are more sensitive than dielectric permittivity in response to the broad phase transition in certain cases. The broad *M-T* phase transition is difficulty to detect by structure analysis method[8-10,12] due to their close related structures near the MPB. Such broad ferroelectric phase transition accompanied by extremely small lattice distortion arise very minor change in polarization, thus, the dielectric response is quite weak. The elastic property and Raman intensity show well-defined anomalies, which reveal a lattice instability due to a soft mode[12]. Future more, it is interesting to note the M-T phase transition drastically shifted to lower temperature with the increase of BT content while Tc remains nearly unchanged, which is attributed to the composition closer to T-stable region with BT increasing and curie temperature of BT being lower than that of PT.

Fig. 3 shows the temperature dependence of piezoelectric $d_{31}$ constant within the temperature range 120-360K. The curves for $d_{31}$ constant for soft PZTs (Navy-type Ⅲ) and hard PZTs (Navy-type Ⅴ) obtained by Zhang *et al*[4] are provided for the purpose of comparison. The $d_{31}$ is calculated according to the formula: $-d_{31} = k_{31}\sqrt{\varepsilon_{33}^T s_{11}^E}$. PMN-BT-PT ceramics exhibit much higher piezoelectric constant than that of hard PZTs. The peak value of piezoelectric constants shifts gradually to lower temperature and broadens with the increase of BT content. Notably, the $d_{31}$ of PMN-BT-PT with compositions of x=2, 3, and 4 show very minor temperature dependence in a wide temperature range. Sample with $x=2$ is found to have the best piezoelectric characteristics, with $d_{31}$ being greater than 170 pC/N from 185 to 360 K and $d_{33}$ of 530 pC/N at 300 K (the inset of Fig. 1(c)). The reduced temperature dependence in PMN-BT-PT ceramics can be understood in terms of the counter-balance between the dielectric and elastic response in the broad phase transition. As shown in Fig. 1, the broad phase transition enhances both the dielectric permittivity and the elastic compliance constant in the observed temperature range, resulting in, naturally, the enhancement of its piezoelectric response. Moreover, for compositions of x=2, 3, and 4, the dielectric constant and elastic compliance show opposite temperature variation in a wide temperature range. The nearly identical in scale but opposite contribution from dielectric and elastic response gives rise to a nearly temperature independent piezoelectric constant in this temperature range.

In summary, we investigated the broad phase *M-T* transition and temperature dependence of piezoelectric properties of PMN-BT-PT ceramics with composition near MPB. The large and temperature-stable piezoelectric response was achieved by a broad phase transition enhanced dielectric and elastic response at lower temperatures. A $d_{31}$ constant as high as 170 pC/N was obtained for the compound with x=2 in a temperature range from 185 to 360 K.

**ACKNOWLEDGEMENTS**

This work is supported by the National Key Project for Basic Research of China (2009CB623305), the National Nature Science Foundation of China (Nos. 61137004 , 51107140 and 50977088), and the External Cooperation Program of the Chinese Academy of Sciences (No. GJHZ1042), and a Grant-in-Aid for Scientific Research, MEXT, Japan (B:22340078).

**FIGURE LEGENDS**

FIG. 1 Dielectric permittivity and dielectric loss (a), elastic compliance constants (b) *vs.* temperature. Composition dependence of Curie temperature at room temperature is shown in the insert (a).

FIG. 2 The intensity ratio of the 750 cm$^{-1}$ peak to 510 cm$^{-1}$ peak for poled samples of *x*=2 (a) and *x*=1 (b). Raman spectra at various temperatures are shown in the insert.

FIG. 3 Temperature dependence of d$_{31}$ constant for PMN-BT-PT ceramics. Composition dependence of d$_{33}$ constant at room temperature is also shown in the insert (c).

# FIGURES

**Fig. 1**

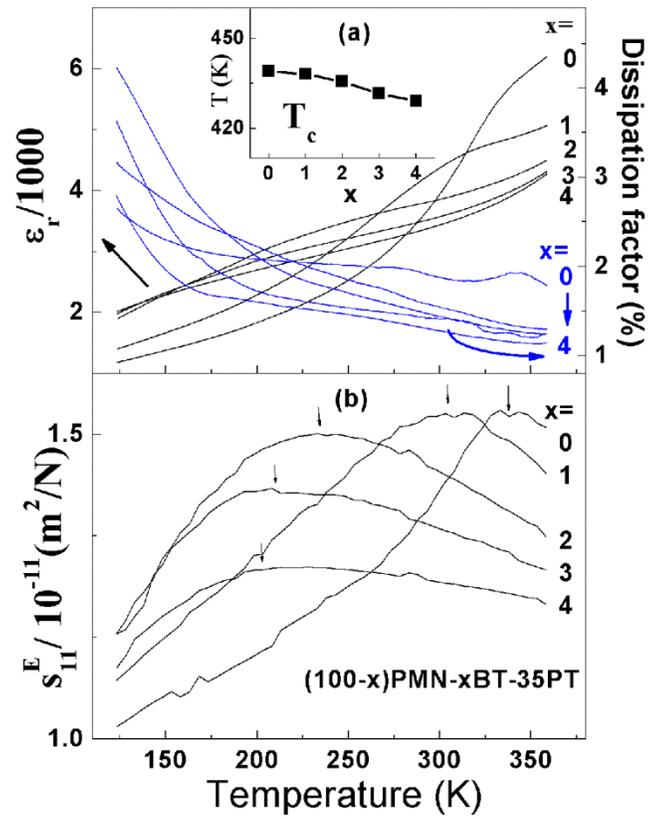

**Fig.2**

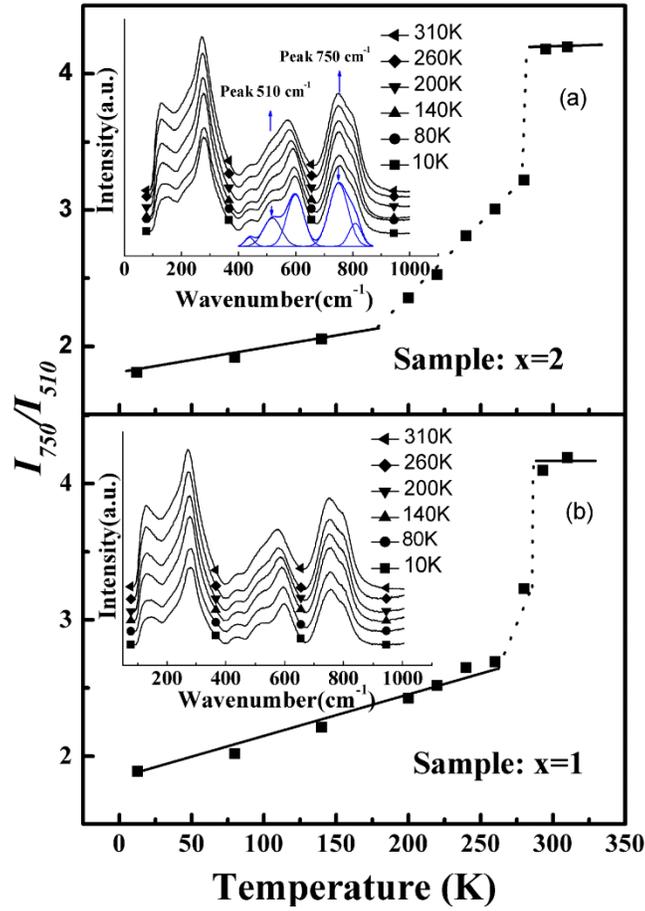

**Fig. 3**

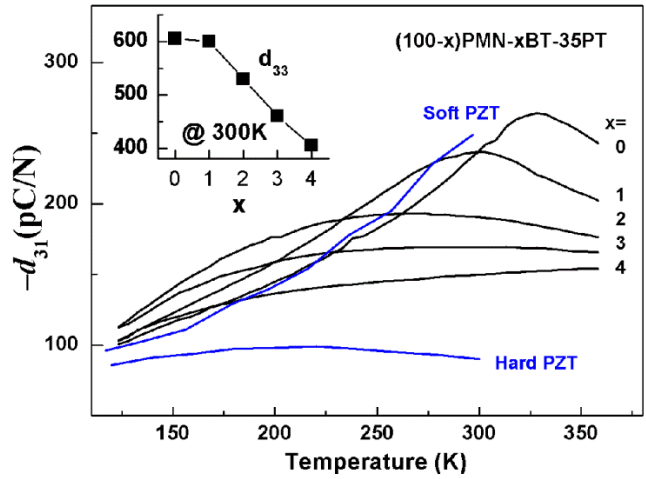